\documentclass[aps,prc,showpacs,floatfix,preprintnumbers,nofootinbib,
amssymb,amsmath]{revtex4}
\usepackage{graphicx,bm,color}
\usepackage{amssymb}
\usepackage{subfigure}
\usepackage{amsmath}
\usepackage{slashed}
\newcommand{\id}{{1\!\!1}}
\def \bsubeq{\begin{subequations}}
\def \esubeq{\end{subequations}}
\def \beq{\begin{equation}}
\def \eeq{\end{equation}}
\def \beqa{\begin{eqnarray}}
\def \eeqa{\end{eqnarray}}

\begin{document}

%%====================================================================%%
\title{Isospin symmetry breaking and baryon-isospin correlations from
Polyakov$-$Nambu$-$Jona-Lasinio model}

\author{Abhijit Bhattacharyya}
\email{abphy@caluniv.ac.in}
\affiliation{Department of Physics, University of Calcutta,
             92, A.P.C. Road, Kolkata - 700009, INDIA}
\author{Sanjay K. Ghosh}
\email{sanjay@jcbose.ac.in}
\author{Anirban Lahiri}
\email{anirbanlahiri.physics@gmail.com}
\author{Sarbani Majumder}
\email{sarbanimajumder@gmail.com}
\author{Sibaji Raha}
\email{sibaji@jcbose.ac.in}
\author{Rajarshi Ray}
\email{rajarshi@jcbose.ac.in}
\affiliation{Center for Astroparticle Physics \&
Space Science, Bose Institute, Block-EN, Sector-V, Bidhan Nagar,
Kolkata-700091, INDIA 
 \\ \& \\ 
Department of Physics, Bose Institute, \\
93/1, A.P.C. Road, Kolkata - 700009, INDIA}
%%====================================================================%%

%%====================================================================%%
\begin {abstract}
We present a study of the 1+1 flavor system of strongly interacting
matter in terms of the Polyakov$-$Nambu$-$Jona-Lasinio model.
We find that though the small isospin symmetry breaking brought
in through unequal light quark masses is too small to affect the
thermodynamics of the system in general, 
it may have significant effect in baryon-isospin correlations and
have a measurable impact in heavy-ion collision
experiments.
\end{abstract}
\maketitle
%%====================================================================%%

%%====================================================================%%
\section{Introduction}
Signatures of phases of matter with deconfined color charges is under
critical investigation for last few decades, both theoretically and
experimentally. Quantum Chromodynamics (QCD) is the formulation for
first principle studies of strongly interacting matter.
Along with the local color symmetry, the quark sector has
few global symmetries also.
In the chiral limit for two light flavors $u$ and $d$,
we have global vector and axial vector symmetry
$SU_V(2) \otimes SU_A(2)$.
For non-zero quark masses, the axial symmetry $SU_A(2)$ is
explicitly broken, while for non-zero quark mass difference
vector (isospin) symmetry $SU_V(2)$
is explicitly broken. 
At low energies the isospin symmetry breaking
(ISB) has relevance in many aspects of hadronic observables
\cite{rusetsky}. Apart from the quark mass difference,
ISB effects may be brought in by electromagnetic
contributions as well. Low energy $\pi-\mathrm{K}$ scattering
has been studied considering the inclusion
of electromagnetic correction into the effective Lagrangian
\cite{nehme_prd65}. In the chiral quark model,
ISB of valence and sea quark distributions in protons and
neutrons has been studied in  \cite{ma_plb,song_prd} and
thermodynamics has been discussed in 
\cite{fraga1}. ISB may also have significant effect in the
context of existence of CP violating phase \cite{creutz_pos}.
Some Lattice QCD investigation of the effect of unequal quark
masses was done in Ref.\cite{gavgup} and recently the effect of
ISB on different hadronic observables were studied in 
Ref.\cite{lattice_rm123,lattice_UKQCD}.
Within the framework of chiral perturbation
theory the isospin breaking effect in quark
condensates has been studied
considering $m_u \neq m_d$ and electromagnetic corrections as well,
where the authors have given an analysis of scalar susceptibilities
\cite{nicola_2011,nicola_2012}.
Both of the above-mentioned effects have been incorporated
also in Nambu$-$Jona-Lasinio (NJL) model \cite{FIK_2007}
to study the influence of the isospin symmetry breaking on
the orientation of chiral symmetry breaking.

In the context of high energy heavy ion collisions where strongly
interacting matter is supposed to exist in a state of thermal and
chemical equilibrium, the ISB effects have not been explored much.
Fluctuations and correlations of conserved charges are important and
sensitive probes for heavy ion physics. Most of the theoretical
studies in this respect are in isospin symmetric limit
(see {\it e.g.} \cite{Gottieb_prl,Gavai_prd2,Alton,Blaizot,purnendu1,
purnendu2,Sasaki_njl,ratti1,RajarshidaA,RajarshidaB,RajarshidaC,fuku2,
pqmsch,anirban1,anirban2,najmul,Sandeep}). Here we present our
first case study of ISB effect on fluctuations and correlations of
strongly interacting matter within the framework of the Polyakov loop
enhanced Nambu$-$Jona-Lasinio (PNJL) model. We discuss the possible
experimental manifestations of the ISB effects based on quite general
considerations in the limit of small current quark masses.
%%====================================================================%%

%%====================================================================%%
\section{Formalism}
 In the last few years PNJL model has appeared in
several forms and context to study the various aspects of phases of
strongly interacting matter (see e.g. \cite{meisinger,fukushima,megias,
ueda,beta,abuki1,abuki2,ohnishi,ratti1,RajarshidaC}).
Here we use the form of the 2 flavor PNJL model with the Lagrangian
as in Ref.\cite{RajarshidaB,RajarshidaC};
\begin{eqnarray}
 \mathcal{L}_{PNJL} = &-&\mathcal{U}[\Phi[A],\bar{\Phi}[A],T]
+\bar{\psi}(\slashed D -\hat{m})\psi \nonumber\\
&+&G_1[(\bar{\psi}\psi)^2+(\bar{\psi}\vec{\tau}\psi)^2
+(\bar{\psi}i \gamma_5 \psi)^2+
(\bar{\psi}i \gamma_5 \vec{\tau}\psi)^2] \nonumber\\
&+&G_2[(\bar{\psi}\psi)^2-(\bar{\psi}\vec{\tau}\psi)^2
-(\bar{\psi}i \gamma_5 \psi)^2+
(\bar{\psi}i \gamma_5 \vec{\tau}\psi)^2]
\end{eqnarray}
$\mathcal{U}[\Phi[A],\bar{\Phi}[A],T]$ is the effective
potential expressed
in terms of traced Polyakov loop and its charge conjugate:
\begin{equation*}
 \Phi=\frac{\mathrm{Tr}L}{N_c}~~~~~~~~~\bar{\Phi}
 =\frac{\mathrm{Tr}{{L}^\dag}}{N_c}
\end{equation*}
Although $\Phi$ and $\bar\Phi$ are complex valued fields, in
the mean field approximation their expectation values are
supposed to be real \cite{Dumitru_prd72}.
In all the previous studies the $u$ and $d$ quarks were considered to
be degenerate. Here we shall consider a mass matrix of the form:
%%####################################################################%%
\begin{align}
\hat{m} & = m_1 \id_{2 \times 2} - m_2 \tau_3 \nonumber \\
& =
\begin{pmatrix}
m_1-m_2 \hfill & 0 \\
0 & m_1+m_2 \hfill 
\end{pmatrix}
\equiv
\begin{pmatrix}
m_u \hfill & 0 \\
0 & m_d \hfill
\end{pmatrix}. \nonumber
\end{align}
%%####################################################################%%
\noindent
where, $\id_{2 \times 2}$ is the identity matrix in flavor space and
$\tau_3$ is the third Pauli matrix. Here $m_u$ and $m_d$ are the current
masses of the $u$ and $d$ quarks respectively. While a non-zero $m_1$
breaks the chiral $SU_A(2)$ symmetry explicitly a non-zero $m_2$ does the
same for the isospin $SU_V(2)$ symmetry. We shall restrict ourselves to
$G_1=G_2=G$ which implies $m_2 = (M_d - M_u)/2$, where $M_u$ and $M_d$ are
the constituent masses of the $u$ and $d$ quarks respectively.

The thermodynamic potential in the mean field approximation is given by,
\begin{eqnarray}
\Omega &=& 2G_1(\sigma^2_u+\sigma^2_d)+4G_2\sigma_u\sigma_d+
{\cal {U^\prime}}[\Phi[A],\bar \Phi[A],T]
-6{\sum_{f=u,d}}{\int_{0}^{\Lambda}}
\frac{d^3p}{(2\pi)^3} E_{f}\nonumber\\
&-& 2T{\sum_{f=u,d}}{\int_0^\infty}\frac{d^3p}{(2\pi)^3}
\ln\left[1+3\Phi e^{-\frac{(E_{f}-\mu_f)}{T}}+3{\bar \Phi}
e^{-2\frac{(E_{f}-\mu_f)}{T}}
+e^{-3\frac{(E_{f}-\mu_f)}{T}}\right]\nonumber\\
&-& 2T{\sum_{f=u,d}}{\int_0^\infty}{\frac{d^3p}{(2\pi)^3}}
\ln\left[1+3{\bar \Phi}e^{-\frac{(E_{f}+\mu_f)}{T}}+3\Phi
e^{-2\frac{(E_{f}+\mu_f)}{T}}+e^{-3\frac{(E_{f}+\mu_f)}{T}}\right]
\end{eqnarray}
where $\sigma_u$ and $\sigma_d$ are the condensates for $u$ and $d$
quarks respectively. Here ${\cal {U^\prime}}[\Phi[A],\bar \Phi[A],T]$
is the modified Polyakov loop potential \cite{RajarshidaC};
\begin{equation*}
\frac{{\cal {U^\prime}}}{T^4}=\frac{{\cal {U}}}{T^4} -
             \kappa \ln J[\Phi,\bar \Phi]
\end{equation*}
where $J[\Phi,\bar \Phi]=\frac{27}{24\pi^2}(1-6{\bar \Phi}\Phi+
4({\bar\Phi}^3+\Phi^3)-3({\bar \Phi}\Phi)^2)$ is known as Vandermonde
determinant and $\kappa$ is a phenomenological parameter, taken to
be 0.1 here.

The behavior of different charge susceptibilities can be studied from
corresponding chemical potential derivatives of the pressure ($P$) obtained
from the thermodynamic potential. In general the $n^{th}$ order diagonal
and off-diagonal susceptibilities are respectively given by, 
${\chi}^{X}_{n}=\dfrac{\partial^n(P/T^4)}{\partial(\mu_X/T)^n}$ and
${\chi}^{XY}_{ij}=\dfrac{{\partial}^{i+j}(P/T^4)}
{\partial(\mu_X/T)^i\partial(\mu_Y/T)^j}$ with $i+j=n$.
These susceptibilities are related to the fluctuations and correlations
of conserved charges $X$ and $Y$ with corresponding chemical potentials
$\mu_X$ and $\mu_Y$. Given a two flavor system the global charge conservation
is expected for baryon number $B$, (third component of) isospin $I_3$ and
electric charge $Q$. In the isospin symmetric limit one can easily
check that the $B-I$ correlation vanishes exactly. For an explicit isospin
symmetry breaking, this correlation may be non-zero. Therefore it
is an interesting and important observable that we wish to study here.

%%====================================================================%%

%%====================================================================%%
\section {Results and Discussions} 

Here we consider the average quark mass $m_1=(m_u+m_d)/2$ fixed at
0.0055 GeV and study the effect of ISB
with three representative values of $m_2=(m_d-m_u)/2$. 
The parameter set in the NJL sector has been determined
separately for the different values of $m_2$ and
the differences in the parameter values were found to
be practically insignificant. The bulk thermodynamic properties
of the system expressed through pressure,
energy density, specific heat, speed of sound etc.\ did not show
significant dependence on $m_2$. Even the diagonal susceptibilities
were almost identical to those at the isospin symmetric limit.  
However, interesting differences were observed for the off-diagonal
susceptibilities in the $B-I$ sector. We first discuss the results
for $\mu_B=0$ and then move to finite $\mu_B$.
%%====================================================================%%

%%====================================================================%%
\subsection{Off-diagonal Susceptibilities for $\mu_B=0$}

%%####################################################################%%
\begin{figure} [!htb]
{\includegraphics [scale=0.5] {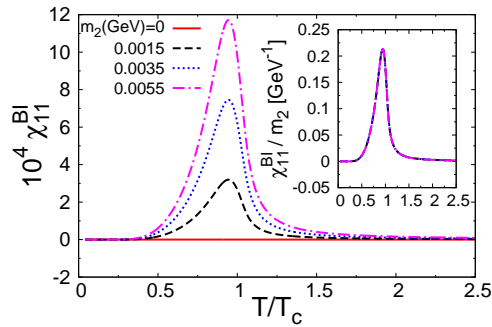}}
\caption{(Color online) Second order off-diagonal
susceptibility in $B-I$ sector at $\mu_B=0$.}
\label{fg.bi_2}
\end{figure}
%%####################################################################%%

In Fig.\ref{fg.bi_2} the second order off-diagonal susceptibility
$\chi^{BI}_{11}$, is plotted against $T/T_c$ for different values
of $m_2$. Here $T_c$ is the crossover temperature obtained from the
inflection point of the scalar order parameters - the mean values
of chiral condensate and Polyakov Loop 
\cite{RajarshidaA,RajarshidaB,RajarshidaC}. As expected we find
$\chi^{BI}_{11}=0$ for $m_2=0$. For non-zero $m_2$ we find
$\chi^{BI}_{11}$ to have non-zero values that change non-monotonically
with the increase in temperature. At low temperatures the excitations
are suppressed due to large constituent masses as well as confining effects
of the Polyakov loop. As the constituent masses and confining effects
decrease with the increase in temperature, the correlations are enhanced.
The peak value appears very close to $T_c$. Thereafter as the constituent
masses become small with respect to the corresponding temperature, the
correlation drops and approaches zero at very high temperatures.

The sensitivity of $\chi^{BI}_{11}$ on $m_2$ is clearly visible. An
exciting feature observed here is that there is an almost linear scaling
of $\chi^{BI}_{11}$ with $m_2$. This is shown in the inset of
Fig.\ref{fg.bi_2}.

%%####################################################################%%
\begin{figure} [!htb]
{\includegraphics [scale=0.5] {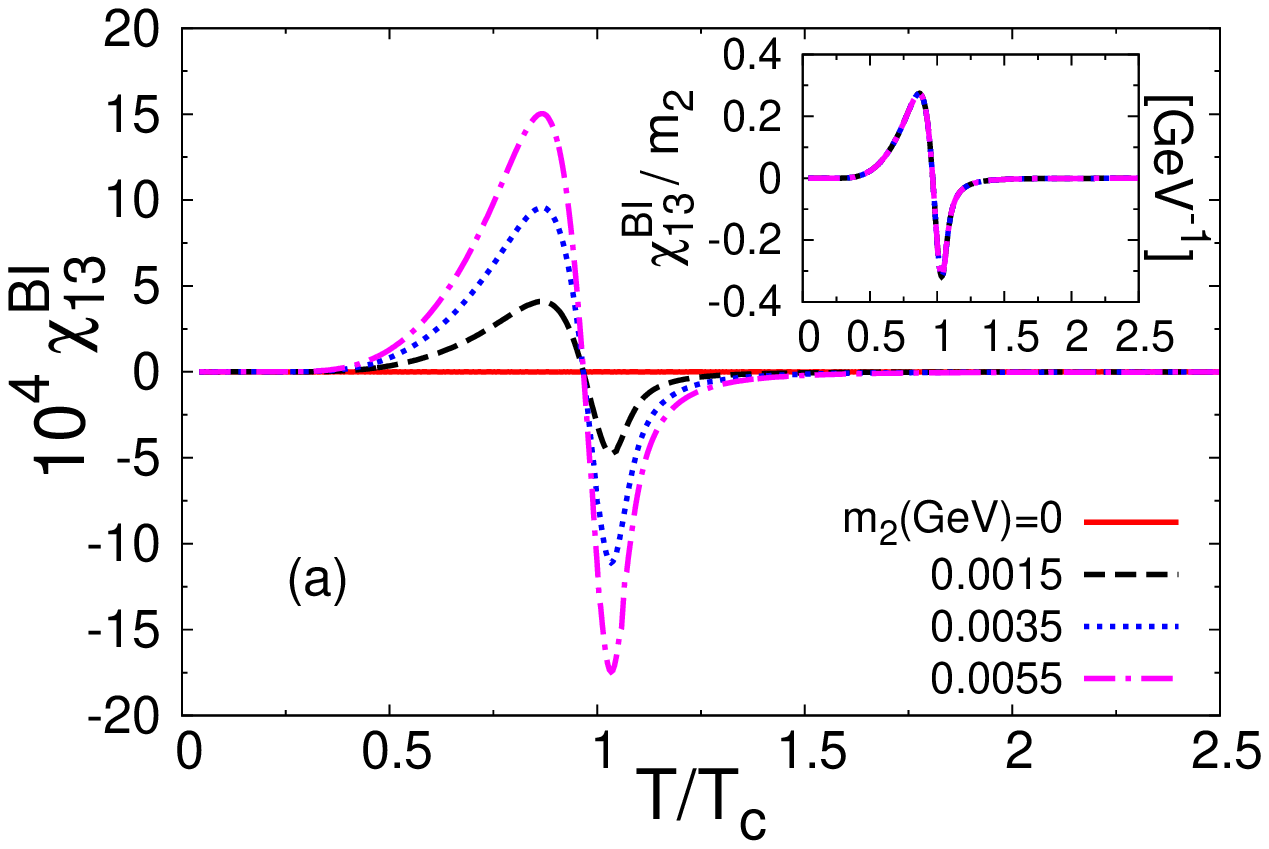} ~~~~
\includegraphics [scale=0.5] {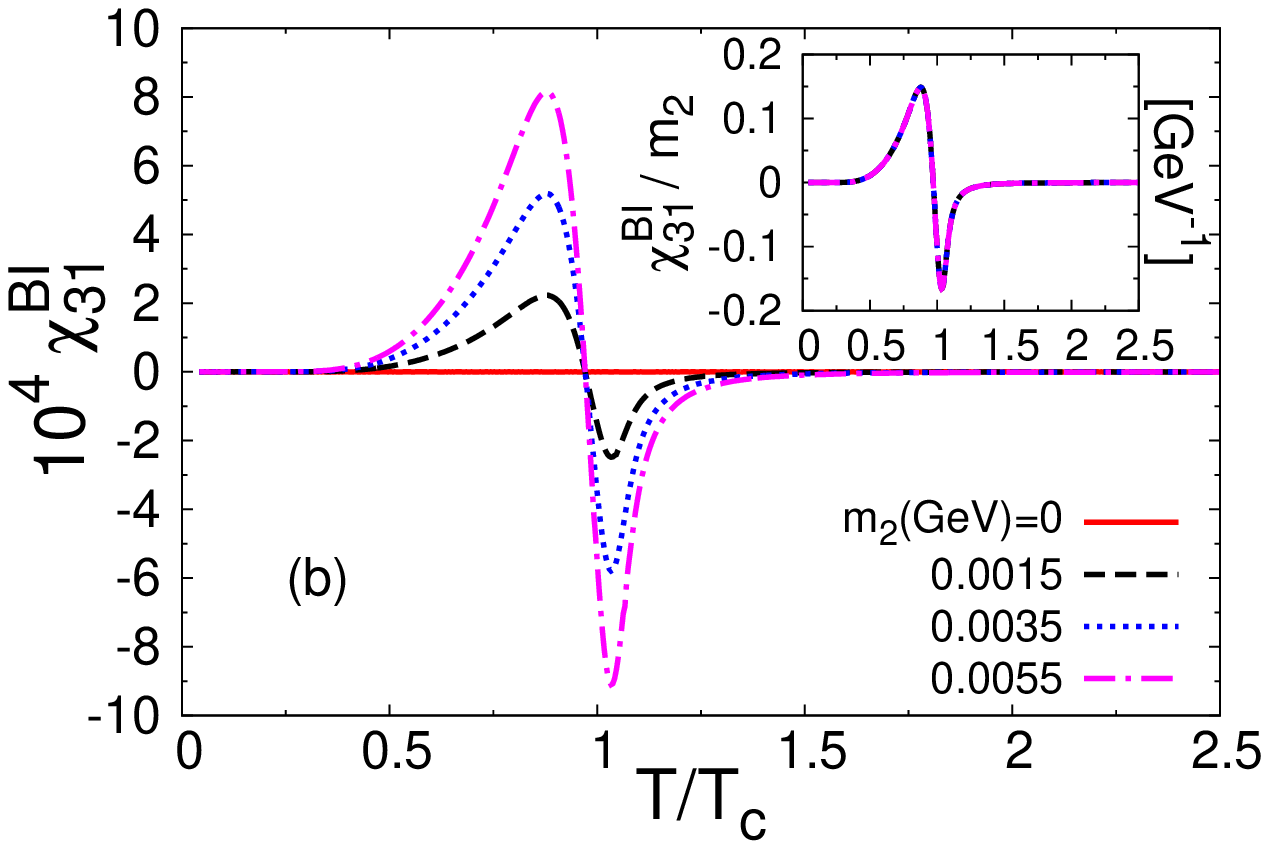}}
\caption{(Color online) Behavior of 4th order off diagonal
susceptibility for different $m_2$.}
\label{fg.bi_4}
\end{figure}
%%####################################################################%%

The fourth order off-diagonal susceptibilities in the $B-I$ sector are 
$\chi_{13}^{BI}$, $\chi_{31}^{BI}$ and $\chi_{22}^{BI}$. The $m_2$ 
dependence of $\chi_{22}^{BI}$ was found to be insignificant. For
$\mu_B=0$, the $T$ dependence for the other two susceptibilities along
with their $m_2$ scaling is shown in Fig.\ref{fg.bi_4}. The qualitative
features of the variation of $\chi_{13}^{BI}$ and $\chi_{31}^{BI}$ with
temperature may be understood by noting that these quantities are
correlators between $\chi_{11}^{BI}$ with those of the isospin
fluctuation $\chi_{2}^{I}$ and the baryon fluctuation $\chi_{2}^{B}$
respectively. In our earlier studies
\cite{RajarshidaA,RajarshidaB,RajarshidaC}, we found that the both
$\chi_{2}^{I}$ and $\chi_{2}^{B}$ increase monotonically with
increasing temperature. On the other hand $\chi_{11}^{BI}$ first
increases up to $T \sim T_c$ and then decreases with increase in $T$,
as shown in Fig.\ref{fg.bi_2}. Therefore one expects that $\chi_{11}^{BI}$
has a positive correlation with $\chi_{2}^{I}$ and $\chi_{2}^{B}$
below $T_c$ and is anti-correlated above $T_c$.

To understand the presence of $m_2$ scaling for some correlators and
absence in others we first note that the different $B-I$ correlators may
be expressed in terms of those in the flavor space. The corresponding
relation between the chemical potentials are
$\mu_u=\frac{1}{3}\mu_B+\frac{1}{2}\mu_I$
and $\mu_d=\frac{1}{3}\mu_B-\frac{1}{2}\mu_I$. This implies,

%%####################################################################%%
\begin{equation}
\chi_{11}^{BI}=\frac{1}{6}({\chi}^u_2-{\chi}^d_2).
\label{eq.chi11BI}
\end{equation}
%%####################################################################%%

The flavor diagonal susceptibilities can be expanded in a Taylor series
of the quark masses around $m_u = m_d = 0$.

%%####################################################################%%
\begin{eqnarray}
\chi_2^f (m_u,m_d) =
\sum_{n=0}^\infty \sum_{i=0}^n a_{i,j}^f m_u^i m_d^j
\label{eq.chi2f_fullexpand}
\end{eqnarray}
%%####################################################################%%

\noindent
where, $a_{i,j}^f=\frac{1}{i! j!}
\Big[\frac{\partial^n \chi_2^f}{\partial m_u^i \partial m_d^j} 
\Big]_{m_u = m_d = 0}$ are the Taylor coefficients,
with $i+j = n$ and $f \in u,d$. Here $a_{0,0}^u$ and
$a_{0,0}^d$ are respectively $u$ and $d$ flavor susceptibilities 
in the chiral limit; hence they are equal.
Moreover, response of $\chi_2^u$ to a change in $m_u$ ($m_d$)
and that of $\chi_2^d$ to a change in $m_d$ ($m_u$) are
identical in the chiral limit.
Thus we have $a_{i,j}^u = a_{j,i}^d$, $\forall~i,j$.
Therefore we get,

%%####################################################################%%
\begin{eqnarray}
&&\chi_2^u (n^{th} {\rm{order}}) - \chi_2^d (n^{th} {\rm{order}})
\nonumber \\
&&= \sum_{i=0}^n \alpha_i m_u^i m_d^i (m_d^{n-2i} - m_u^{n-2i}).
\label{eq.diff}
\end{eqnarray}
%%####################################################################%%

\noindent
where $\alpha_i=a_{i,n-i}^u = a_{n-i,i}^d$. It is clear that for
any given $n$ and $i$, the R.H.S.\ contains a factor $(m_d-m_u)$.
Therefore $\chi_{11}^{BI}$ (Eq.\ref{eq.chi11BI}) is proportional
to $m_2$ if the terms for $n\geq 3$ are sub-dominant in
Eq.\ref{eq.chi2f_fullexpand}. This is what we observed for the
range of $m_2$ considered here.

For the higher order correlators one can similarly write,

%%####################################################################%%
\begin{eqnarray}
\chi_{13}^{BI}&=&\frac{1}{24}({\chi}^u_4-{\chi}^d_4
+2\chi_{13}^{ud}-2\chi_{31}^{ud}) \label{eq.bi_13}\\
\chi_{31}^{BI}&=&\frac{1}{54}({\chi}^u_4-{\chi}^d_4
-2\chi_{13}^{ud}+2\chi_{31}^{ud}) \label{eq.bi_31} \\
\chi_{22}^{BI}&=&\frac{1}{36} (\chi_4^u+\chi_4^d-
2\chi_{22}^{ud}) \label{eq.bi_22}
\end{eqnarray}
%%####################################################################%%

\noindent
For all these quantities the first two terms on R.H.S.\ were found
to be dominant. Considering again the Taylor expansion of $\chi_4^f$
(but obviously with Taylor coefficients different from that of $\chi_2^f$)
in quark masses, $\chi_{13}^{BI}$ and $\chi_{31}^{BI}$ were found to be
proportional to $m_2$. Since $\chi_{22}^{BI}$ contains sum of fourth
order flavor fluctuations instead of their difference, no $m_2$ scaling
appeared in this case.
%%====================================================================%%

%%====================================================================%%
\subsection{Off diagonal Susceptibilities for $\mu_B\neq 0$}

%%###################################################################%%
\begin{figure} [!htb]
{\includegraphics [scale=0.47] {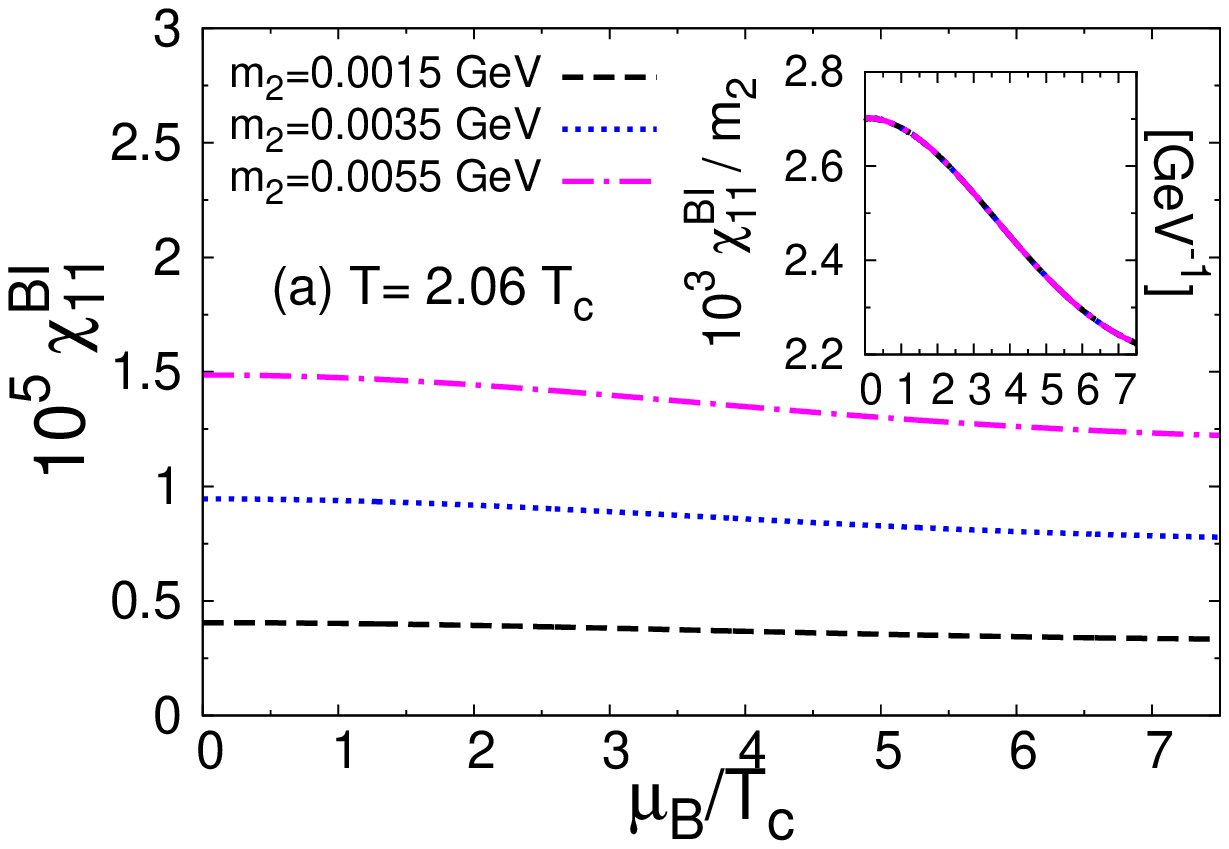} ~~~~
\includegraphics [scale=0.47] {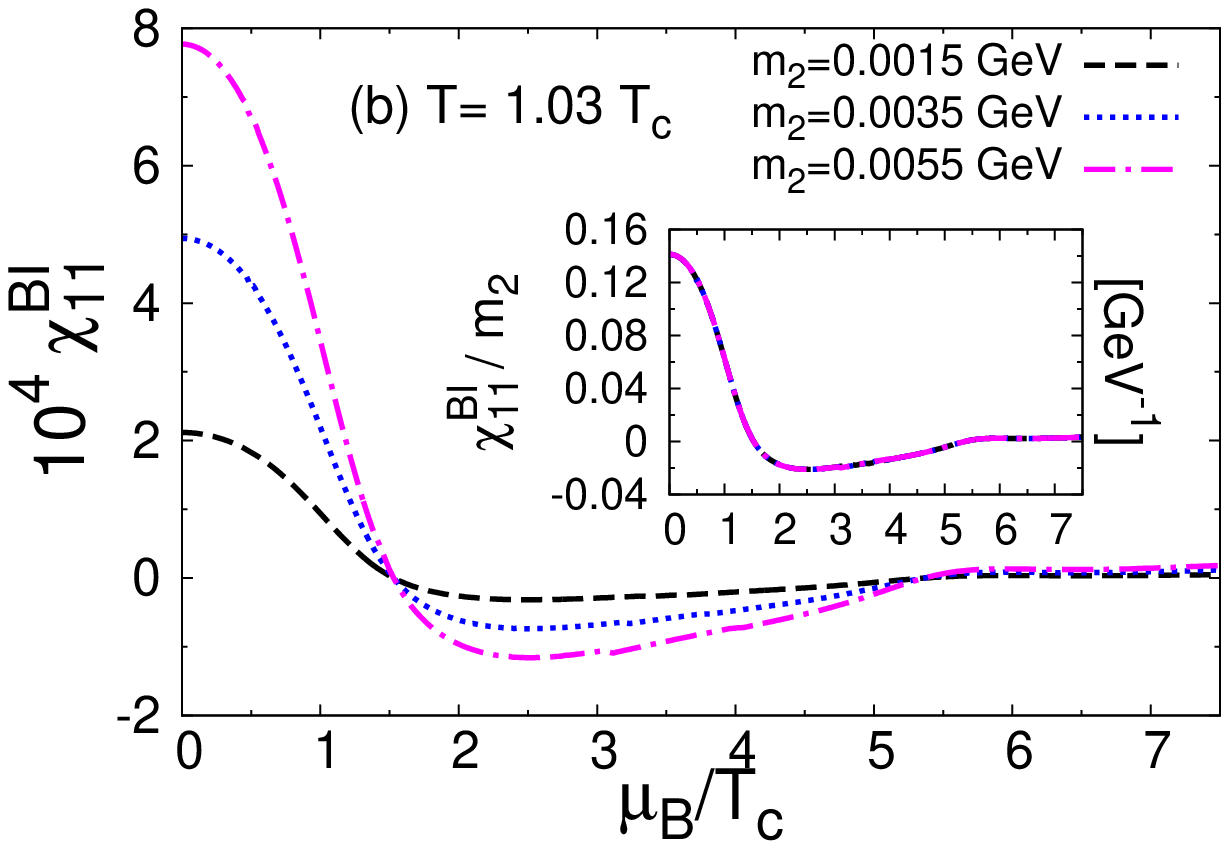} \\ \vspace{0.5cm}
\includegraphics [scale=0.47] {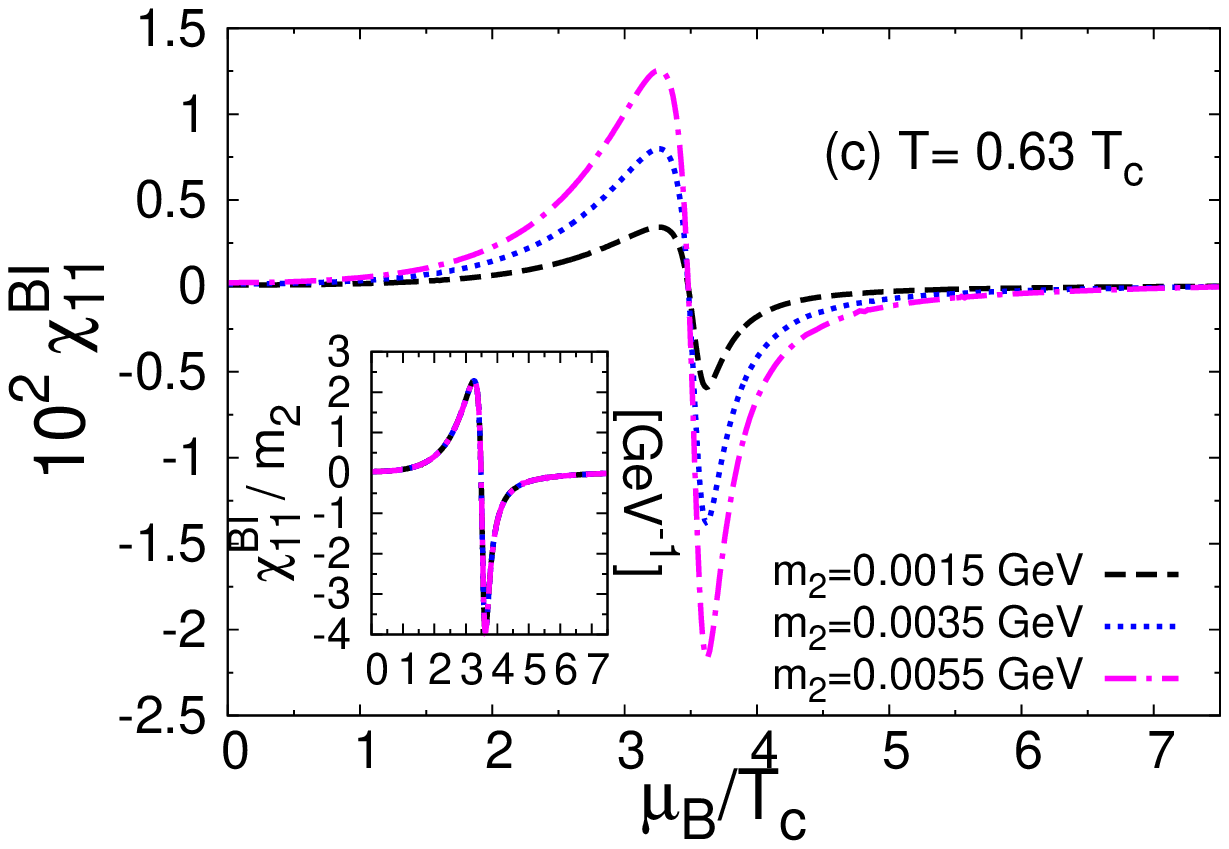} ~~~~
\includegraphics [scale=0.47] {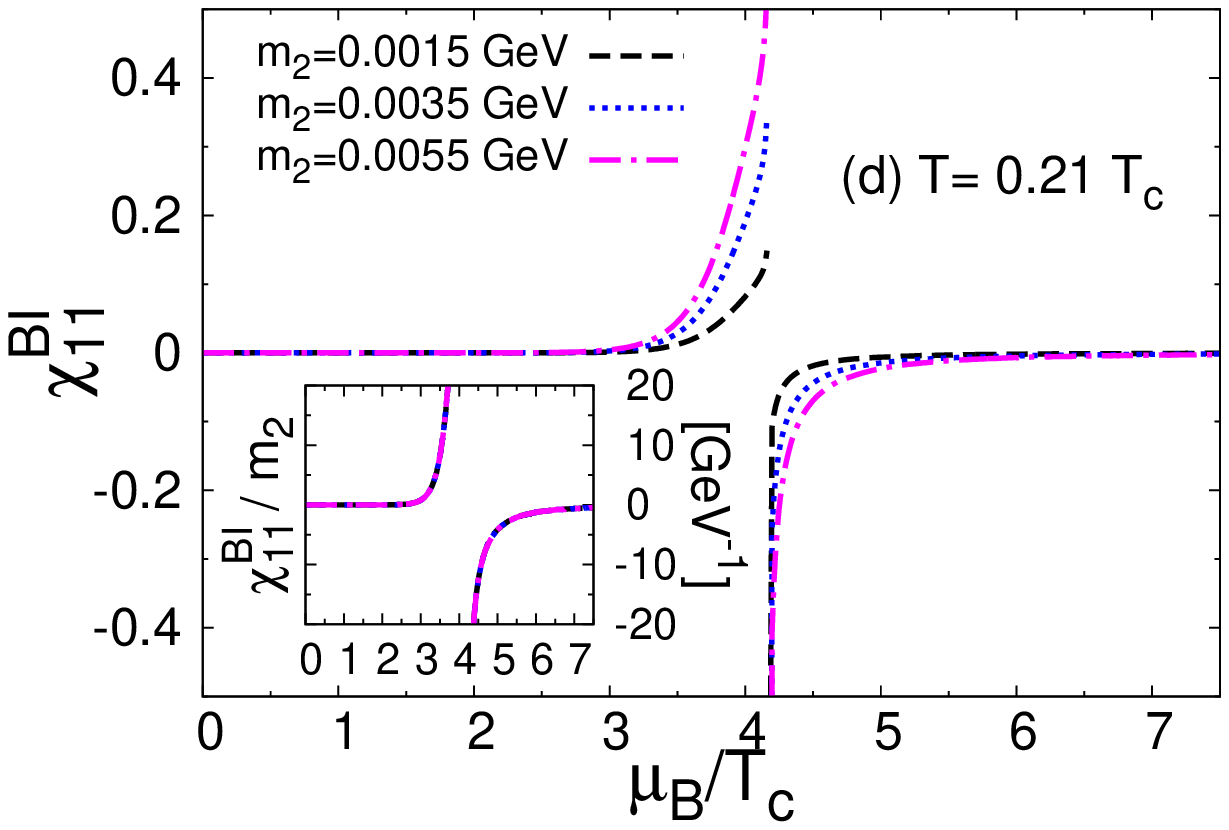}}
\caption{(Color online) $\chi^{BI}_{11}$
along baryon chemical potential at different temperatures.}
\label{fg.bi_11_mu}
\end{figure}
%%###################################################################%%

In Fig.\ref{fg.bi_11_mu}, the variation of $\chi^{BI}_{11}$ with $\mu_B$
is shown for four different temperatures. The features vary widely over
the different ranges of temperature and chemical potential. At
$T \sim 2 T_c$, $\chi^{BI}_{11}$ is positive, and slowly decreases
with increasing $\mu_B$. Close to $T_c$, $\chi^{BI}_{11}$ drops sharply
to zero, becomes negative and then again slowly approaches zero. Going
down somewhat below $T_c$ there is an initial increase in
$\chi^{BI}_{11}$ for some range of $\mu_B$, and thereafter it follows
the behavior at $T_c$. Finally at very low temperatures the change in sign of
$\chi^{BI}_{11}$ is marked by a discontinuity, arising due to a first
order phase boundary which exists in this range of $T$ and $\mu_B$.

%%###################################################################%%
\begin{figure}  [!htb]
 {\includegraphics [scale=0.45] {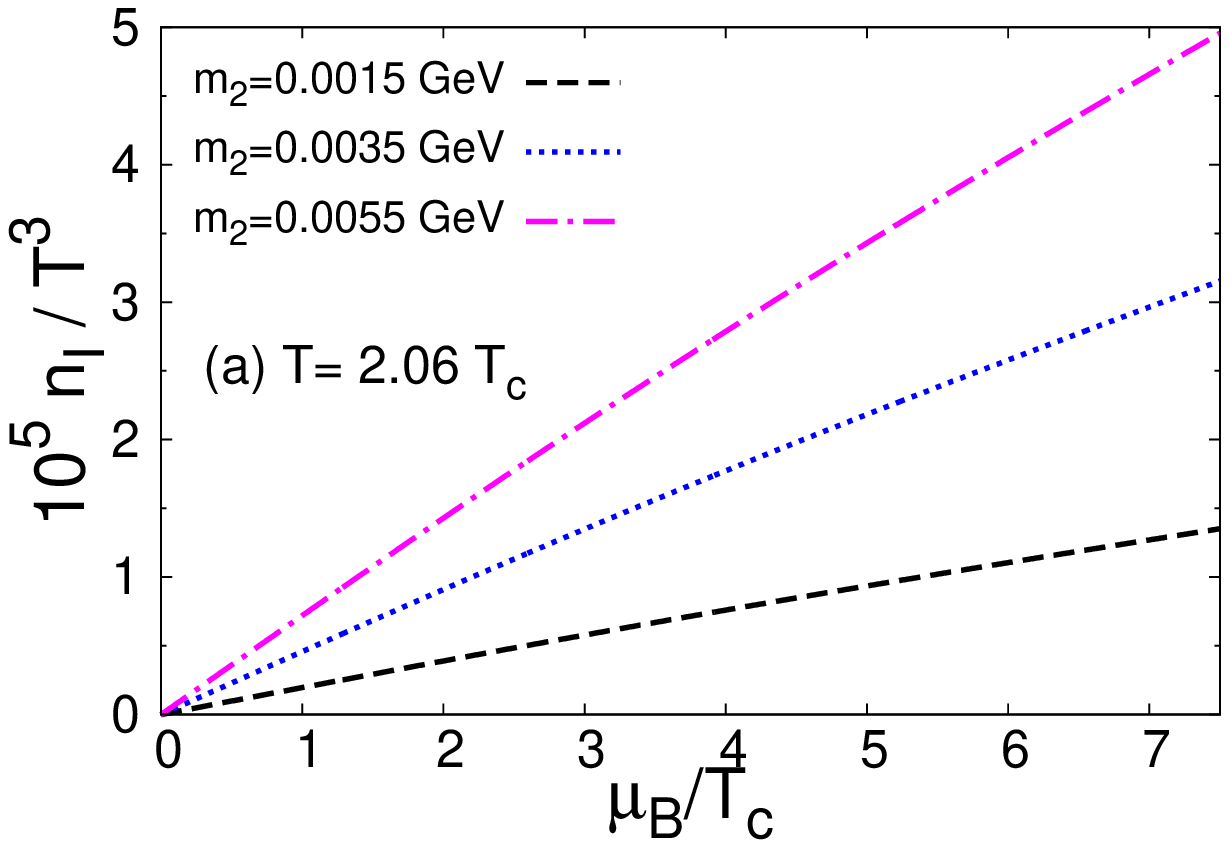} ~~~~
\includegraphics [scale=0.45] {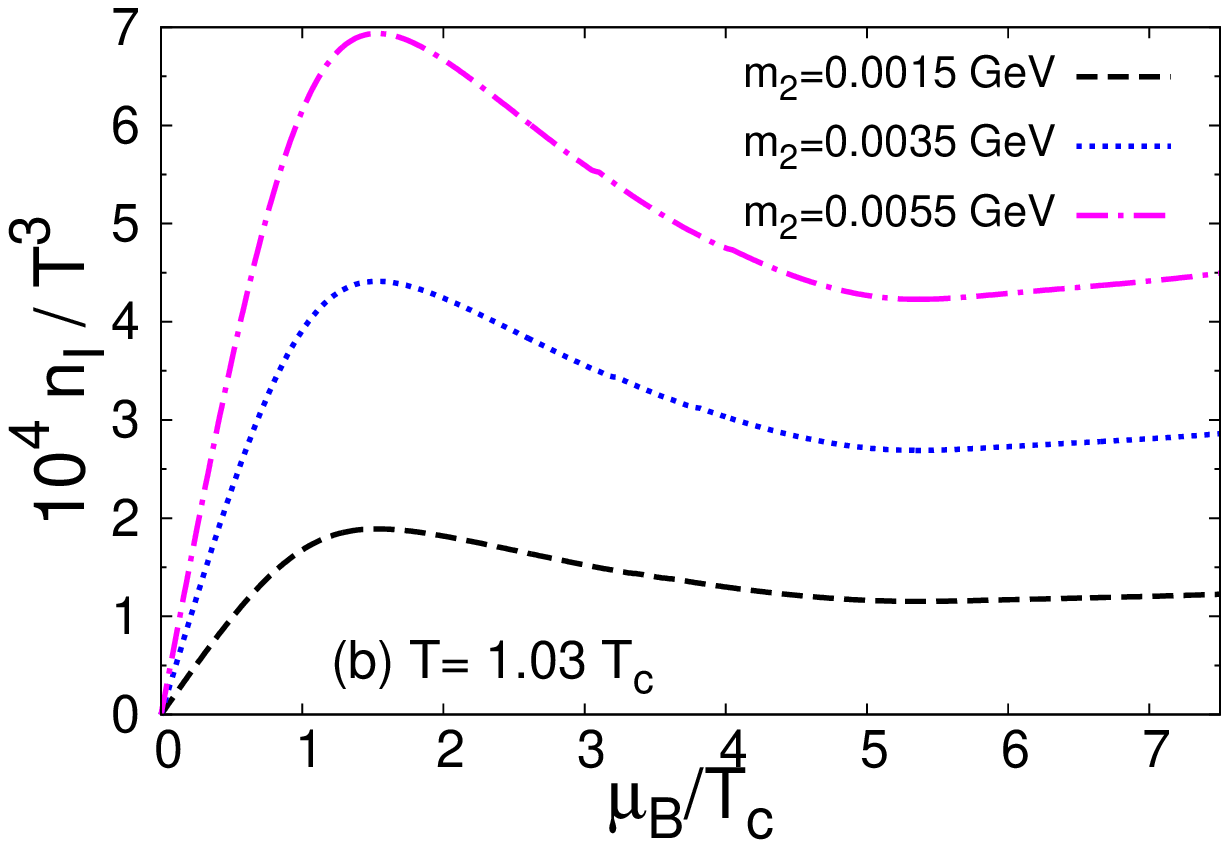} \\ \vspace{0.5cm}
\includegraphics [scale=0.45] {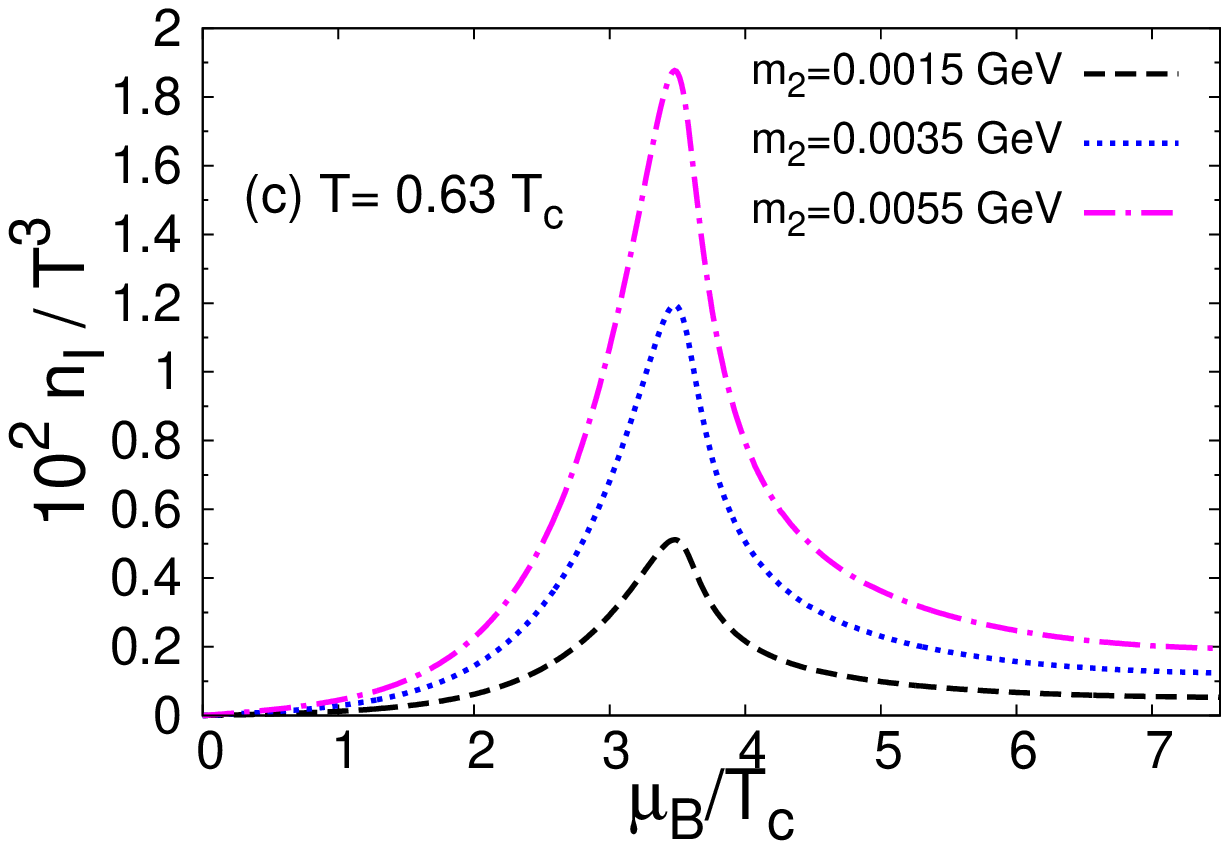} ~~~~
\includegraphics [scale=0.45] {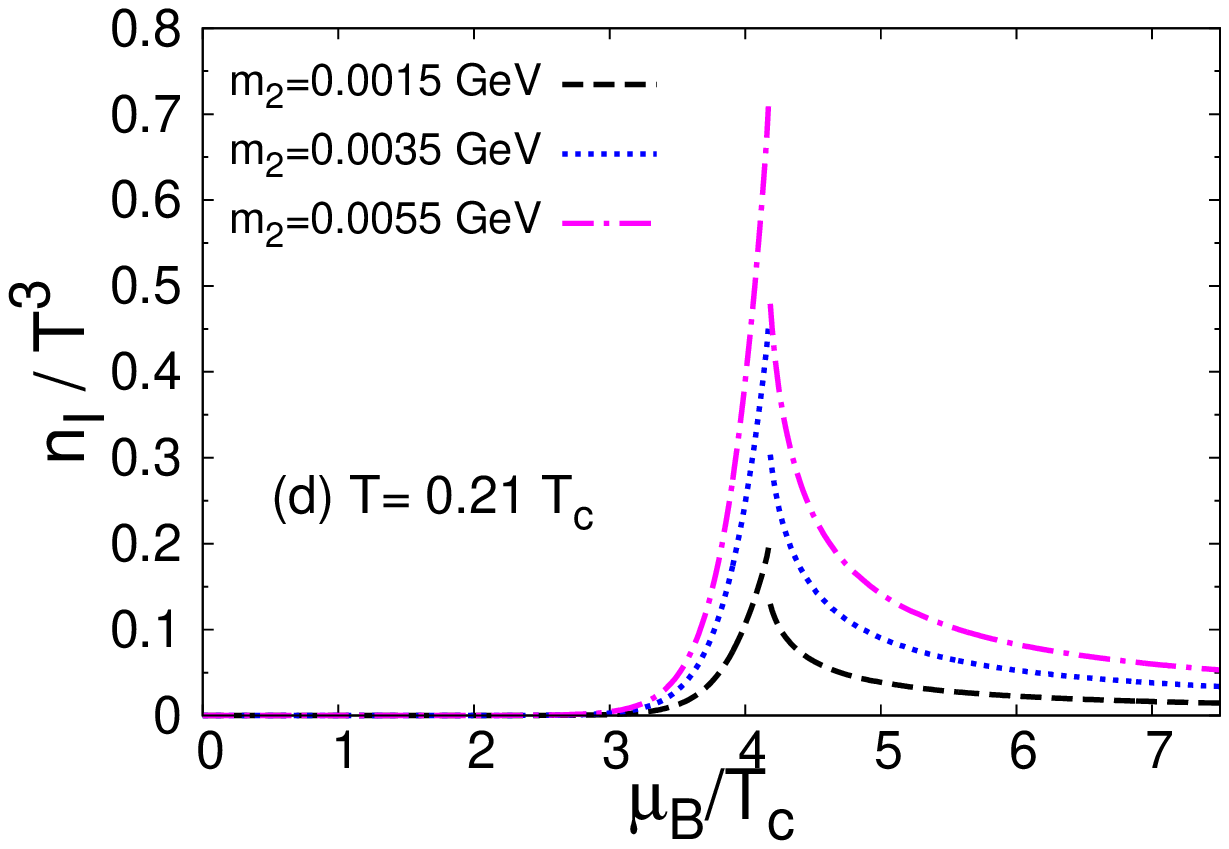}}
\caption{(Color online) Isospin number density along $\mu_B$
at the same temperatures as in Fig.\ref{fg.bi_11_mu}.}
\label{fg.inum}
\end{figure}
%%###################################################################%%

These various features can be understood by expressing
$\chi^{BI}_{11}= \frac{\partial}{\partial \mu_B}(\frac{\partial P}
{\partial \mu_I})=(\frac{\partial n_I}{\partial \mu_B})$, where $n_I$
is the isospin number density. It is worth noting that although we have
considered $\mu_I=0$ throughout the present study, a non-zero isospin
number is generated due to non-vanishing $m_2$. So let us study the
behavior of isospin number density with changing baryon chemical
potential. Now $n_I = (n_u - n_d)/2$, where $n_u$ and $n_d$ are respectively
the $u$ and $d$ quark densities. The number density of a given flavor
at a constant temperature is governed by the corresponding mass as well
as the chemical potential. The isospin number should be positive as the
$u$ quark mass is smaller than that of $d$ quark. With increase in
baryon chemical potential this difference is expected to increase giving
an increasing $n_I$. This expected feature is found to hold in the low
temperatures for a range of $\mu_B$ as can be seen from Fig.\ref{fg.inum}.
However there is a subsequent drop in isospin number as the constituent
quark masses suddenly start to fall beyond a critical $\mu_B$, gradually
becoming insignificant as the constituent masses reduce to the current
mass values. The rise and fall of $n_I$ explains the complete behavior
of $\chi^{BI}_{11}$ for $T<T_c$. In fact the same explanation applies
for the other two temperatures in the following way. Close to $T_c$
the constituent masses of the quarks are again approaching the current
mass values. $n_I$ is increasing with $\mu_B$, but too slowly and
therefore $\chi^{BI}_{11}$, given by the slope, starts dropping. The
latter part still follows the behavior of $T < T_c$. By $2 T_c$ the
current mass is almost achieved and $n_I$ increases almost linearly
with a very small slope with respect to $\mu_B$. The corresponding
$\chi^{BI}_{11}$ is positive and decreasing very slowly.

An amazing fact remains that the scaling of the correlators with $m_2$
survives for all conditions of $T$ and $\mu_B$. This is shown in the
insets of Fig.\ref{fg.bi_11_mu}. A major implication is that all higher
order derivatives of $n_I$ with respect to $\mu_B$ would also show
similar scaling behavior. This can be seen by expanding $\chi^{BI}_{11}$
in a Taylor series in $\mu_B$ about $\mu_B=0$ as,

%%####################################################################%%
\begin{eqnarray}
{\chi}^{BI}_{11}(\mu_B)={\chi}^{BI}_{11}(0)+
\frac{{\mu}_B^2}{2!}{\chi}^{BI}_{31}(0)
+\frac{{\mu}_B^4}{4!}{\chi}^{BI}_{51}(0)+
\cdot \cdot \cdot \cdot \cdot
\label{eq.bi_11_mu}
\end{eqnarray}
%%####################################################################%%

\noindent
In the above series odd order terms vanish due to $CP$ symmetry. Since
${\chi}^{BI}_{11}(\mu_B)$ on the L.H.S.\ scales with $m_2$, the same can
be expected to hold true individually for all the coefficients on the
R.H.S.\ up to any arbitrary order. The first two Taylor coefficients have
already been shown to follow the scaling relation in Fig.\ref{fg.bi_2} and
Fig.\ref{fg.bi_4}(right panel) respectively.
%%====================================================================%%

%%====================================================================%%
\subsection {Further implications of ISB in Heavy Ion Collisions}
%%####################################################################%%
\begin{figure} [!htb]
{ \includegraphics [scale=0.5] {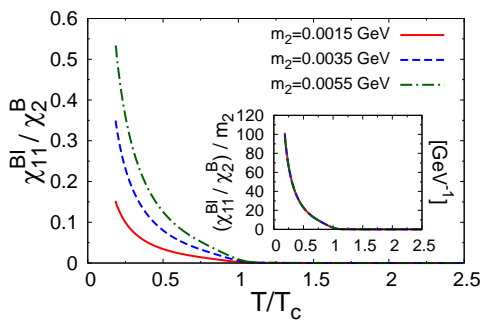} }
\caption{(Color online) Ratio of B$-$I correlation to baryon number
fluctuation at $\mu_B=0$.}
\label{fg.ratio}
\end{figure}
%%####################################################################%%

%%####################################################################%%
\begin{figure} [!htb]
{\includegraphics [scale=0.40] {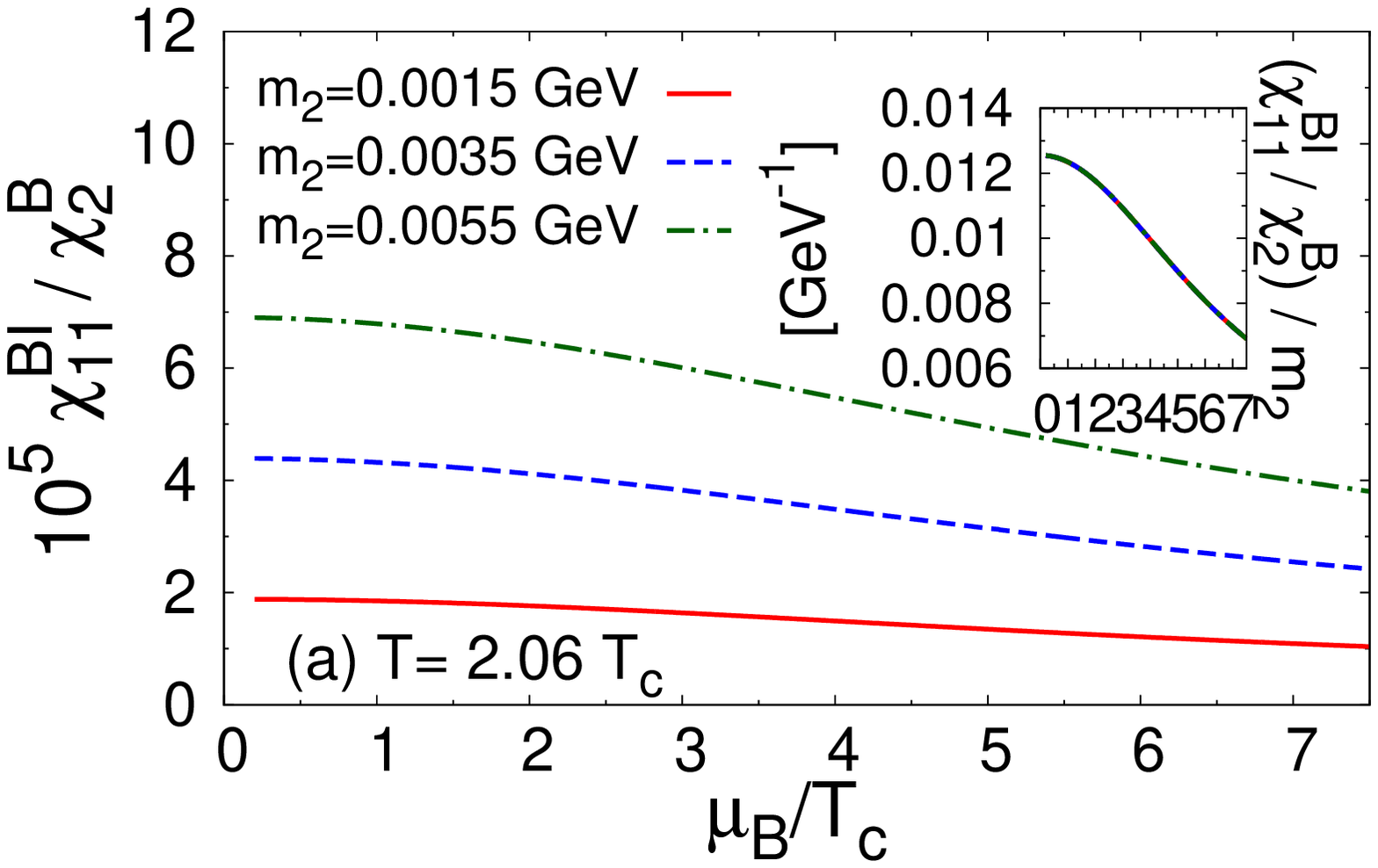} ~~~~
\includegraphics [scale=0.40] {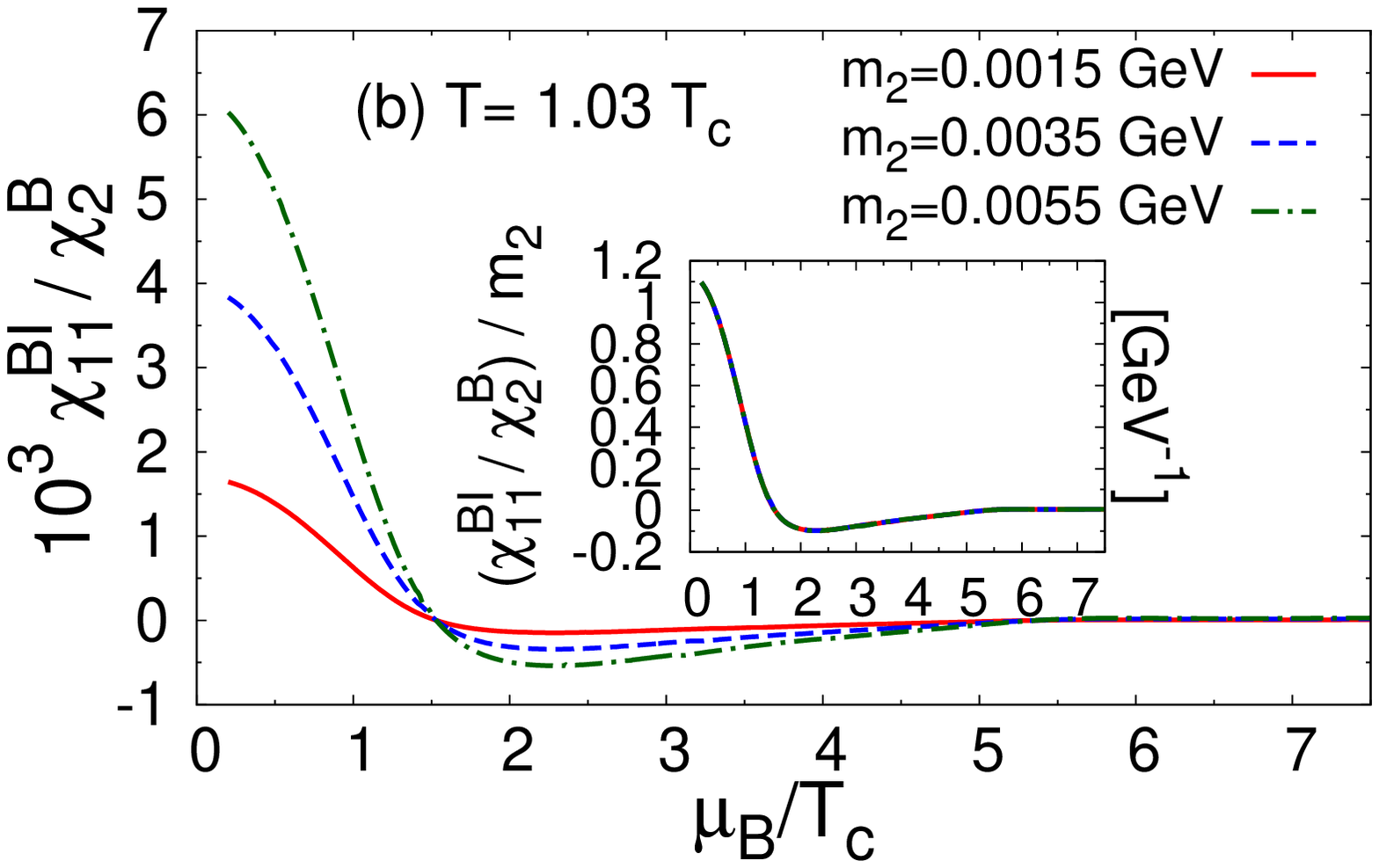}\\ \vspace{0.5cm}
\includegraphics [scale=0.40] {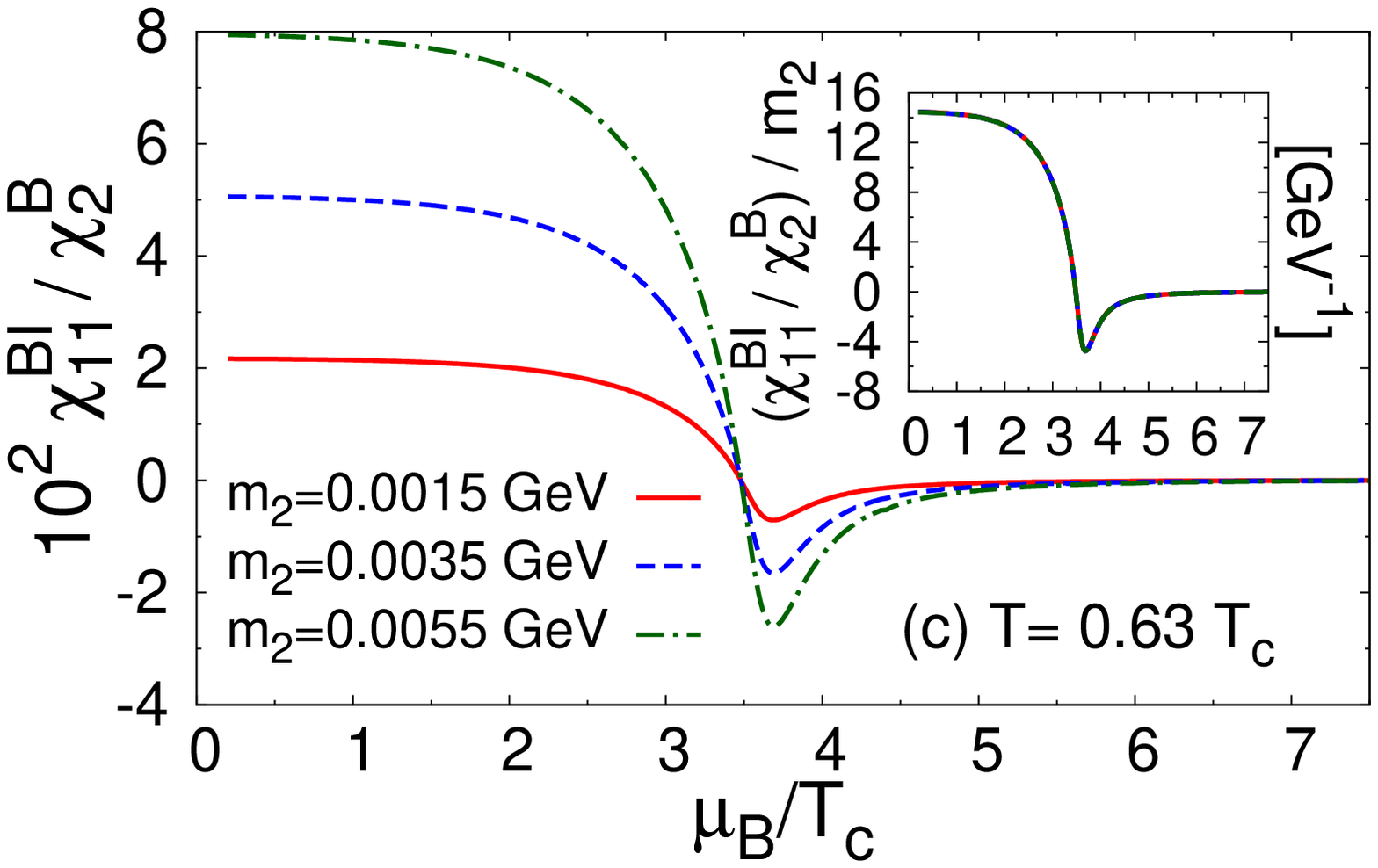} ~~~~
\includegraphics [scale=0.40] {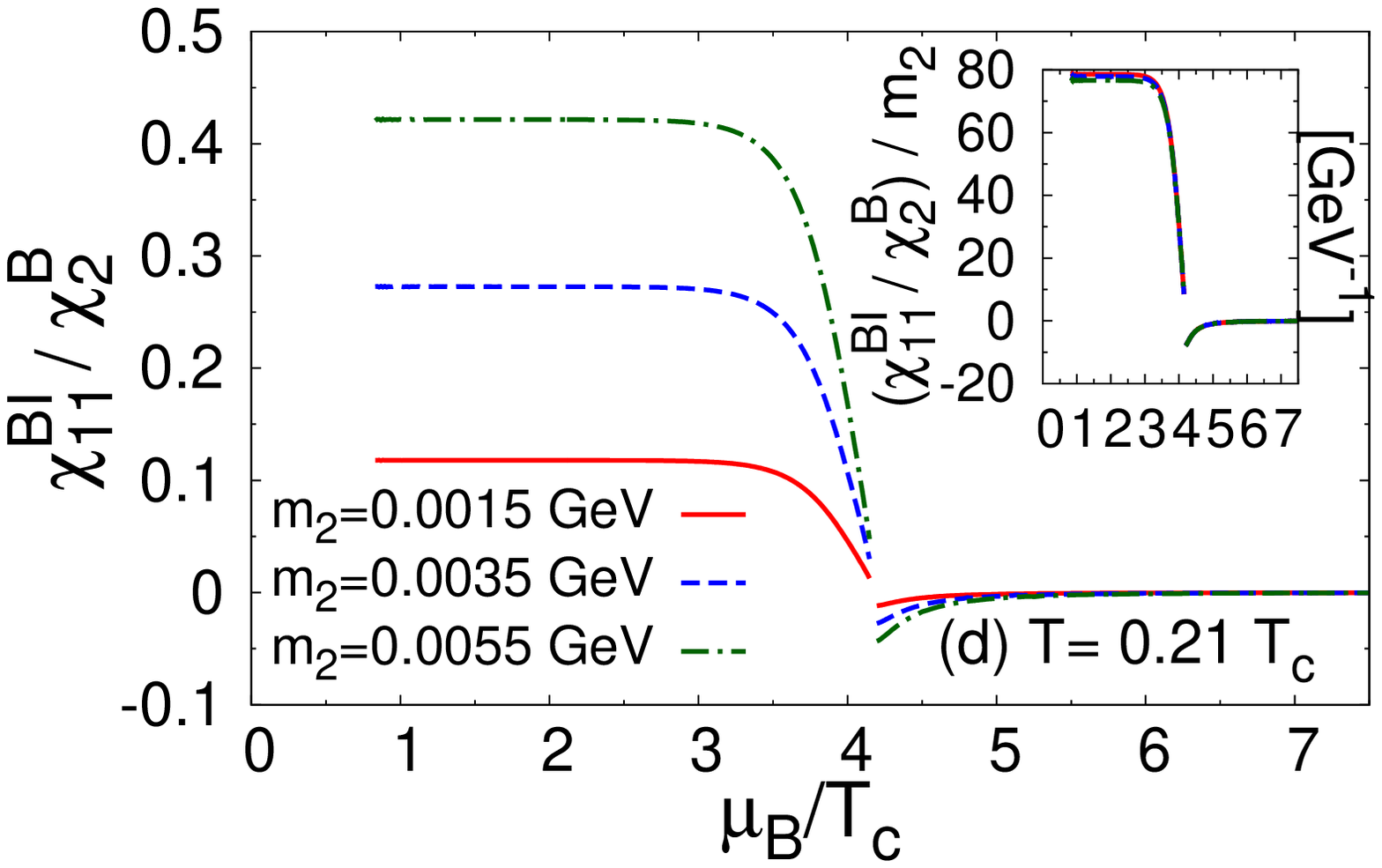}}
\caption{(Color online) Ratio of B$-$I correlation to baryon number
fluctuation along $\mu_B$ at the same temperatures as in
Fig.\ref{fg.bi_11_mu}.}
\label{fg.ratio-mub}
\end{figure}
%%####################################################################%%

Correlation between conserved charges, is an experimentally measurable
quantity obtained from event-by-event analysis in heavy-ion collisions
\cite{kochreview}. To compare with experiments it is often useful to
consider ratios such as ${\rm R}_2=\chi_{11}^{BI}/\chi_{2}^{B}=
\mathrm{C}_{BI}/\mathrm{C}_{BB}$ \cite{kochreview,agarwal}. Here
$ \mathrm{C}_{XY}=\frac{1}{N_E}\sum_{i=1}^{N_E} X_iY_i-
(\frac{1}{N_E}\sum_{i=1}^{N_E}X_i)\cdot
(\frac{1}{N_E}\sum_{i=1}^{N_E}Y_i), $
where $N_E$ is the total number of events considered and $X_i$ and $Y_i$
are the event variables corresponding to the conserved charges in a given
event $i$. Ratios of this kind are practically useful in eliminating
uncertainties in the estimates of the measured volume of the fireball.
Relevance of similar ratios of fluctuations have also been discussed in
the Lattice QCD framework \cite{sgupta,bazarov}.

The temperature variation of ${\rm R}_2$ obtained here is shown in 
Fig.\ref{fg.ratio}. It decreases monotonically and approaches zero above
$T_c$. This is expected as the baryon number fluctuation increases much
more rapidly than the $B-I$ correlation below $T_c$, and thereafter
$\chi^{BI}_{11}$ goes to zero while $\chi^B_2$ attains a non-zero value.

Though not completely monotonic, ${\rm R}_2$ goes down to extremely
small values close to the phase/crossover boundary for $\mu_B \ne 0$.
This is shown in Fig. \ref{fg.ratio-mub}. If freeze-out of the particles
produced in heavy-ion collisions occurs very close to phase/crossover
boundary after the system has passed through the partonic phase then
$R_2$ will have very small values. A systematic study of this ratio can
thus indicate how close one could approach the phase boundary in
heavy-ion collisions. In fact a small negative value of ${\rm R}_2$
for intermediate energy experiments where the temperature is supposed
to be quite low would be an exciting indicator of a phase transition.

The $m_2$ scaling that we observed for $\chi^{BI}_{11}$ or ${\rm R}_2$
is most likely model independent as it is expected on very general
grounds for small current quark masses as discussed above. Therefore,
at any temperature and chemical potential, one can use the $m_2$ scaling
to estimate the mass asymmetry of constituent fermions in a physical
system as,

%%###################################################################%%
\begin{equation}
{m_2}^{\mathrm{expt}}=\frac{{\rm R}_2^{\rm expt}(T,\mu_B)}
{{\rm R}_2^{\rm th}(T,\mu_B)}\times {m_2}^{\rm th}
\end{equation}
%%###################################################################%%

\noindent
where, \textquoteleft expt\textquoteright\ and
\textquoteleft th\textquoteright\ denotes the experimentally
measured and theoretically calculated values of the corresponding
quantities respectively. To the best of our knowledge this is the
first theoretical attempt which indicates that quark mass asymmetry
in thermodynamic equilibrium can be directly measured from
heavy-ion collision experiments.

For the fourth order correlators, an important point to note
is that for fractional baryon number of the constituents,
$|\chi_{13}^{BI}| / |\chi_{31}^{BI}| > 1$. It is easy to check
that the inequality is reversed for integral baryon number
{\it i.e.}\ for protons and neutrons. However from
Fig.\ref{fg.bi_4} we see that the former inequality persists
well below $T_c$. This may well be an artifact of the PNJL model.
Therefore it would be in principle interesting to crosscheck the
corresponding results from Lattice QCD. Enhanced statistics of
present and future experiments may make it possible to measure
this extremely sensitive probe. The direction of the above
inequality would be important in deciding if partonic matter
may have been produced in the medium.

We expect that the measurement of these correlations in experiments
pose a big challenge. Firstly, at low temperatures where ${\rm R}_2$
is large, both the numerator and denominator are quite small, making the
measurement difficult. Experimentally these fluctuations are measured
from the cumulants of the multiplicity distribution at chemical freezeout
\cite{STAR_2014}. For example, around highest RHIC energy, particle
ratios are expected to be frozen at $T\sim 0.170$ GeV and $\mu_B \sim$
0.020 GeV and for those values of temperature and baryon chemical
potential $\chi^B_2$ has been measured very accurately. To measure
$\chi^{BI}_{11}$ at same level of accuracy, assuming normally distributed
population, naively the statistics needs to be increased by a factor of
$10^6$ w.r.t.\ the same for existing calculation in case of $\chi^B_2$
which seems to be difficult at the present stage. In view of this
the situation is somewhat better at say $\surd s_{\rm NN} \sim$ 8
GeV, where the freezeout is expected for $T\sim 0.140$ GeV and
$\mu_B \sim$ 0.420 GeV. In this case absolute values of both the
numerator and denominator of ${\rm R}_2$ are well within the
measurable regime for future experimental facilities like BES-II
in RHIC and CBM in FAIR which will have fairly high statistics for
low energy runs and possibly can overcome this problem.

The other experimental challenge is the detection and measurement of
neutrons which along with the protons are supposed to be the highest
contributor to the baryon-isospin correlations. Incidentally the Large
Area Neutron Detector facility has already been developed at GSI,
Darmstadt, Germany, where one can measure neutron properties in heavy-ion
collision experiments up to an incident energy of 1 GeV per nucleon 
\cite{land1,land2,land3}. Hopefully with further developments in
detection technology, relevant data for neutrons may be available for
higher collision energies in near future. 

At this point it seems relevant to mention that the issue of neutron
detection has arisen earlier even for the measurement of baryon
fluctuation itself. In case of baryon number cumulants, methods are given
\cite{KA_C85,KA_C86,BK_C86,BKS_C87} to reconstruct and estimate the
effect of unobserved neutrons as well as other effects like finite
acceptance and global conservation of baryon number. In
Ref.\cite{KA_C85,KA_C86} the key ingredient is the observation
that due to some late stage processes the isospins of different
nucleon species are almost uncorrelated which makes it possible to
write the actual baryon number cumulants in terms of the observed
proton number cumulants. It is argued that for low values of
$\surd s_{\rm NN}$ this randomization of isospin is not favored
and neutron and proton number distribution will not be factorized
in the final state due to the existence of primordial isospin
correlation. This is precisely what we observe in our framework.
From Fig.\ref{fg.bi_2} and Fig.\ref{fg.bi_11_mu} it can be easily seen
that as we go down by $\surd s_{\rm NN}$ ({\it i.e.}\ decreasing $T$
and increasing $\mu_B$), correlation through isospin between different
baryon species, {\it i.e.}\ $\chi^{BI}_{11}$ will increase. Therefore
a direct measurement of neutrons is desirable even for measuring
baryon number fluctuations at low energies apart from the baryon-isospin
correlations. 

Another question that still remains is whether the isospin asymmetry
brought in through nonzero electric charge may disturb the scaling
and inclusion of this QED effect will be another complete study
in itself and will be reported later.
%%====================================================================%%

%%====================================================================%%
\section{Conclusion}
In the present paper we have investigated the effect of isospin symmetry
breaking through the unequal masses of $u$ and $d$ flavor. The work is
done within the framework of 1+1 flavor PNJL model. The main result found
is the observation that the off-diagonal susceptibilities in the $B-I$
sector depend on a small current quark-mass difference, whereas
the bulk thermodynamic properties of the system (pressure, energy density,
specific heat, speed of sound) do not show such dependence. The relevance
of conserved charge fluctuations to the study of the transition region of
strongly interacting matter is unquestionable. We showed that the $B-I$
correlations may give important information of the state of matter created
in the heavy-ion collision experiments. Whereas the correlation remains
positive for small $\mu_B$, it may become negative in the high $\mu_B$
partonic phase. The change of sign of the correlation seems to be
completely model independent.

Also the typical scaling behavior of these correlations with the quark
mass difference $m_2$ has been argued to be model independent as long as
the current masses are small. This scaling may enable one to estimate the
quark mass asymmetry in heavy-ion experiments.

Another model independent observable that we discussed is for the
fourth order correlators. Depending on whether the ratio
$|\chi_{13}^{BI}| / |\chi_{31}^{BI}|$ is greater than 1 or not, one may
infer if a partonic phase has been created and survived till freeze-out
in the heavy-ion experiments. In our study the ratio was always found
to be greater than 1, which may be model artifact.

A physically more realistic scenario of course requires the
incorporation of strange quarks and/or QED effects which will be
reported elsewhere.

The experimental observation of baryon-isospin correlation is a
challenging job. Hopefully the future relativistic heavy-ion experiments
with appropriate neutron detectors and high statistics data would be able
to address this important issue.
%%====================================================================%%

%%====================================================================%%
\begin{acknowledgments}
We would like to thank Sourendu Gupta, Rajiv Gavai, Supriya Das,
Paramita Deb, Swagato Mukherjee and Tapan Nayak for many useful
discussions. A.L.\ and S.M.\ thank IFCC (Bose Institute), A.B.\ thanks
UGC (UPE \& DRS) \& DST and R.R.\ thanks DST for financial support.
\end{acknowledgments}
%%====================================================================%%

%%====================================================================%%

%%====================================================================%%

\end{document}